\documentclass[12pt]{article}

\begin{document}
\renewcommand{\thefootnote}{\fnsymbol{footnote}}

\begin{titlepage}
\hfill{hep-th/0210158}
\vspace{15mm}
\baselineskip 8mm

\begin{center}
{\LARGE \bf Solvable ${\cal N}$=(4,4) Type IIA String Theory \\ 
in Plane-Wave Background and D-Branes}
\end{center}
\baselineskip 6mm
\vspace{10mm}
\begin{center}
Seungjoon Hyun$^a$\footnote{\tt hyun@phya.yonsei.ac.kr}
and
Hyeonjoon Shin$^b$\footnote{\tt hshin@newton.skku.ac.kr} \\[5mm]
{\it
$^a$ Institute of Physics and Applied Physics, Yonsei University,
Seoul 120-749, Korea \\
$^b$ BK 21 Physics Research Division and Institute of Basic Science \\
Sungkyunkwan University, Suwon 440-746, Korea}
\end{center}

\thispagestyle{empty}


\vfill

\begin{center}
{\bf Abstract}
\end{center}
\noindent 
We study various aspects of ${\cal N}$=(4,4) type IIA GS superstring theory
in the pp-wave background, which arises as the compactification of maximally
supersymmetric eleven-dimensional pp-wave geometry along the spacelike
isometry direction. We show the supersymmetry algebra of ${\cal N}$=(4,4)
worldsheet supersymmetry as well as non-linearly realized supersymmetry.
We also give quantization of closed string and open string incorporating 
various boundary conditions. From the open string boundary conditions, we
find configurations of D-branes which preserve half the supersymmetries. 
Among these we identify D4 brane configurations with longitudinal five brane
configurations in matrix model on the eleven-dimensional pp-wave geometry.

\vspace{20mm}
\end{titlepage}

\baselineskip 6.5mm
\renewcommand{\thefootnote}{\arabic{footnote}}
\setcounter{footnote}{0}


\section{Introduction and Summary}
Recently, there have been lots of interests in the M theory on 
eleven-dimensional 
pp-wave geometry \cite{kow194}-\cite{bla081} which is maximally supersymmetric 
by admitting 32 Killing
spinors. One way to study the M theory in this background is to use matrix
model \cite{ber021}, which seems to be particularly suitable, 
recalling the light-like nature 
of the background. Matrix model on pp-wave background
\cite{ber021,das185} has interesting property such as the removal of
flat directions because of bosonic mass terms and therefore the moduli
space reduces to the set of isolated points which represent fuzzy spheres.
This matrix model has time dependent supersymmetry and thus  the bosons and
fermions have different masses. 
Using this model various kinds of nonperturbative BPS 
states have been identified \cite{das185}-\cite{par161}. Other aspects of 
the model have been discussed in \cite{kim061}-\cite{mic204}. 

Another way to study the M theory is to compactify the theory along
the circle under which it becomes type IIA superstring theory, whose
coupling is proportional to the compactification radius with fixed
string tension.  Many characteristic features, like nonperturbative
BPS states, of IIA superstring theory is inherited from M theory. In
turn, the study of IIA superstring theory gives some informations on
`mother' M theory.  The eleven-dimensional maximally supersymmetric
pp-wave geometry has various spacelike isometry, along which the M
theory can be compactified\cite{mic140}.

One such choice of the circle direction leads to the following
ten-dimensional pp-wave background\cite{hyu074,sug029}:\footnote{This
geometry comes from the circle compactification of the maximally
supersymmetric eleven dimensional pp-wave.  The direction of
compactification is the isometry $\partial / \partial x^9$, which is
the combination of the original $x^4$ and $x^9$ coordinates.  This is
why we have the non-vanishing two form RR field strength.  For
details, see the appendix of the paper \cite{hyu074}.}
\begin{eqnarray}
& & ds^2 = - 2 dx^+ dx^-
    - A(x^I) (dx^+)^2  + \sum^8_{I=1} (dx^I)^2~,
                                      \nonumber \\
& & F_{+123} = \mu~,\ \ \  F_{+4} = -\frac{\mu}{3}~,
\label{pp-wave}
\end{eqnarray}
where
\begin{equation}
A(x^I) = \sum^4_{i=1} \frac{\mu^2}{9} (x^i)^2
            +\sum^8_{i'=5} \frac{\mu^2}{36} (x^{i'})^2~.
\end{equation}
In contrast to the eleven-dimensional pp-wave geometry, this geometry admits
only 24 Killing spinors. In \cite{hyu074} the full 
type IIA GS superstring action on this ten-dimensional pp-wave 
was constructed and their worldsheet 
supersymmetry has been identified as ${\cal N}=(4,4)$. 
We would like to note that IIA GS string theories on different pp-wave
geometry have been considered in \cite{ali037,cve082,cve229}.

In this paper we continue the study of the (4,4) IIA superstring
theory constructed in \cite{hyu074}.\footnote{The same model was
considered in \cite{sug029} in which the action was modified following
our previous paper \cite{hyu074}. They considered the quantization of
string coordinates which overlap with some materials in sections 2 and
4 in this paper. They also considered the classification of
supersymmetric D-branes, which is, however, different from ours.  In
particular, we have got different results in D4 and D6 branes. As will
be seen in section 4, the two form RR field strength should be taken
care properly in the classification of D-branes.}  In what follows, we
recall, the light-cone gauge fixed, IIA GS superstring action on
pp-wave geometry, given in \cite{hyu074}.  It is very complicate to
get the full expression of the GS superstring action in the general
background (see, for example, \cite{cve202,hyu247}).  However, in the
case at hand, we can use the fact that eleven-dimensional pp-wave
geometry can be thought as a special limit of $AdS_4 \times S^7$
geometry on which the full supermembrane action is constructed using
coset method \cite{dew209}.  The full IIA GS superstring action on
this geometry can be obtained by the double dimensional reduction
\cite{duf70} of the supermembrane action of \cite{dew209} in the
pp-wave limit.  It turns out that, by fixing the fermionic
$\kappa$-symmetry as
\begin{equation}
\Gamma^+ \theta = 0~,
\label{kfix}
\end{equation}
the action reduces to the following form:
\begin{eqnarray}
S_{IIA} 
  &=& -\frac{1}{4 \pi \alpha'} \int d^2 \sigma \sqrt{-h} h^{mn}
  \bigg[ -2 \partial_m X^+ \partial_n X^- 
    + \partial_m X^I \partial_n X^I 
    - A(X^I) \partial_m X^+ \partial_n X^-
                         \nonumber \\
  & & -2 \partial_m X^+ \bar{\theta} \Gamma^- \partial_n \theta
      + \frac{\mu}{2} \partial_m X^+ \partial_n X^+
       \bar{\theta} \Gamma^- 
         \left( \Gamma^{123} + \frac{1}{3} \Gamma^{49}
         \right) \theta
  \bigg]    
                         \nonumber \\
  & &  - \frac{1}{2 \pi \alpha'} \int d^2 \sigma
   \epsilon^{mn} \partial_m X^+ \bar{\theta} \Gamma^{-9} 
    \partial_n \theta ~,
\end{eqnarray}
in which the Majorana fermion $\theta$ is the combination of
Majorana-Weyl fermions $\theta^1$ and $\theta^2$ with opposite ten
dimeniosional chiralities, that is, $\theta = \theta^1 + \theta^2$,
and the fermion part is simply given by quadratic term.  In fact,
recently it has been argued in \cite{miz043,rus114} that for a wide
class of pp-wave background, the fermionic part of GS action has only
quadratic terms.

The equation of motion for $X^+$ is harmonic, the same as in the flat
case, which allows the usual light-cone gauge fixing,
\begin{equation}
X^+ = \alpha' p^+  \tau~,
\end{equation} 
where $p^+$ is the total momentum conjugate to $X^-$. The worldsheet
diffeomorphism can be consistently fixed as $\sqrt{-h} = 1 ~, \ \ \
h_{\sigma \tau} = 0 ~, $ which fix other worldsheet metric components
consistently as $h_{\tau \tau} = -1$ and $ h_{\sigma \sigma} = 1 $.
 
After rescaling the fermionic coordinate as $\theta \rightarrow
\theta / \sqrt{2 \alpha' p^+}$, the light-cone gauge fixed 
action of IIA string is 
given by
\begin{eqnarray}
S_{LC} 
 &=& - \frac{1}{4 \pi \alpha'} \int  d^2 \sigma
 \Bigg[ \eta^{mn} \partial_m X^I \partial_n X^I 
      + \frac{m^2}{9} (X^i)^2
      + \frac{m^2}{36} (X^{i'})^2
                       \nonumber \\
 & & + \bar{\theta} \Gamma^- \partial_\tau \theta 
     + \bar{\theta} \Gamma^{-9} \partial_\sigma \theta 
     - \frac{m}{4} \bar{\theta} \Gamma^- 
        \left( \Gamma^{123} + \frac{1}{3} \Gamma^{49} \right)
        \theta  
  \Bigg] ~,
\end{eqnarray}
where 
\begin{equation}
m\equiv \mu \alpha' p^+
\end{equation}
is a mass parameter which characterizes the masses of the worldsheet
fields.  Therefore the light-cone gauge-fixed action $S_{LC}$ is
quadratic in bosonic as well as fermionic fields and thus describes a
free theory much the same as in the IIB string theory \cite{met044} on
the pp-wave geometry \cite{bla242}. This makes accessible to study IIA
string theory on pp-wave geometry which also helps to understand the M
theory on pp-wave geometry.

Different characteristic feature of IIA string theory on pp-wave
geometry compared to IIB case is the structure of worldsheet
supersymmetry.  Sixteen spacetime supersymmetries which satisfy
$\Gamma^+ \epsilon = 0$ are non-linearly realized on the worldsheet
action.  As is typical in light-cone GS superstring, the remaining
eight spacetime supersymmetries, combined with appropriate kappa
transformations, turn into worldsheet (4,4) supersymmetry of
Yang-Mills type\cite{hyu074}.  Furthermore, in contrast to the matrix
model in eleven dimensions, this, so-called, dynamical supersymmetry
is time independent. Note that, recently in \cite{cve193},
linearly-realized worldsheet supersymmetry of GS superstring in more
general background was discussed.

In order to see all these more 
clearly, we rewrite the action $S_{LC}$  in the 16 component
spinor notation with $\theta^A = \frac{1}{2^{1/4}} \left(
\begin{array}{c} 0 \\ \psi^A
\end{array} \right)$ (Superscript $A$ denotes the $SO(1,9)$
chirality and takes values of 1 and 2.  1 is for the positive
chirality and 2 for the negative one.), under which it becomes
\begin{eqnarray}
S_{LC}
 &=&  - \frac{1}{4 \pi \alpha'} \int  d^2 \sigma
 \Bigg[ \eta^{mn} \partial_m X^I \partial_n X^I 
      + \frac{m^2}{9} (X^i)^2
      + \frac{m^2}{36} (X^{i'})^2
                       \nonumber \\
 & & - i \psi_+^1 \partial_+  \psi^1_+
     - i \psi_-^1 \partial_+  \psi^1_-
     - i \psi^2_+ \partial_- \psi^2_+
     - i \psi^2_- \partial_- \psi^2_-
     +2i \frac{m}{3} \psi^2_+ \gamma^4 \psi^1_-
     - 2i \frac{m}{6} \psi^2_- \gamma^4 \psi^1_+
 \Bigg]~,
                       \nonumber \\
\label{lc-action}
\end{eqnarray}
where $\partial_\pm =\partial_\tau\pm \partial_\sigma$.  Here the sign
of subscript in $\psi^A_\pm$ represents the eigenvalue of
$\gamma^{1234}$.  In our convention, fermion has the same $SO(1,9)$
and $SO(8)$ chirality measured by $\Gamma^9$ and $\gamma^9$,
respectively.

Thus, among sixteen fermionic components in total, eight with
$\gamma^{12349}=1$ have the mass of $m/6$ and the other eight with
$\gamma^{12349}=-1$ the mass of $m/3$, which are identical with the
masses of bosons. Therefore the theory contains two supermultiplets
$(X^i, \psi^1_-, \psi^2_+)$ and $(X^{i'}, \psi^1_+, \psi^2_-)$ of
worldsheet (4,4) supersymmetry with the masses $m/3$ and $m/6$,
respectively.

This can be seen explicitly from the transformation laws \cite{hyu074} 
for the worldsheet (4,4) supersymmetry which are given by
\begin{eqnarray}
\delta X^i &=& \frac{i}{\sqrt{\alpha' p^+}} (
                \psi^1_- \gamma^i \epsilon^1_+  +
                \psi^2_+ \gamma^i \epsilon^2_-)~,   
                                             \nonumber \\
\delta \psi^1_- &=&
  \frac{1}{\sqrt{\alpha' p^+}} 
    \left( \partial_- X^i \gamma^i \epsilon^1_+
        - \frac{m}{3} X^i \gamma^4 \gamma^i 
          \epsilon^2_- 
    \right) ~,
                                                \nonumber \\
\delta \psi^2_+ &=&
  \frac{1}{\sqrt{\alpha' p^+}} 
    \left( \partial_+ X^i \gamma^i \epsilon^2_-
        + \frac{m}{3} X^i \gamma^4 \gamma^i 
           \epsilon^1_+ 
    \right)~,
\label{dynst1}
\end{eqnarray}
for the supermultiplet $(X^i, \psi^1_-, \psi^2_+)$ and
\begin{eqnarray}
\delta X^{i'} &=& \frac{i}{\sqrt{\alpha' p^+}} (
                \psi^1_+ \gamma^{i'} \epsilon^1_+ + 
                \psi^2_- \gamma^{i'} \epsilon^2_- )~,   
                                             \nonumber \\
\delta \psi^1_+ &=&
  \frac{1}{\sqrt{\alpha' p^+}} 
    \left( 
      \partial_- X^{i'} \gamma^{i'} \epsilon^1_+
      +\frac{m}{6} X^{i'} \gamma^4 \gamma^{i'} 
       \epsilon^2_-
    \right) ~,
                                                \nonumber \\
\delta \psi^2_- &=&
  \frac{1}{\sqrt{\alpha' p^+}} 
    \left( \partial_+ X^{i'} \gamma^{i'} \epsilon^2_-
       - \frac{m}{6} X^{i'} \gamma^4 \gamma^{i'}
       \epsilon^1_+ 
    \right) ~,
\label{dynst2}
\end{eqnarray}
for the supermultiplet $(X^{i'}, \psi^1_+, \psi^2_-)$.  As chirality
structure of parameter for dynamical supersymmetry, we note that
$\gamma^{12349} \epsilon^1_+ = - \epsilon^1_+$ and $\gamma^{12349}
\epsilon^2_- = -\epsilon^2_-$.

Under the kinematical supersymmetry, the bosons do not transform:
\begin{equation}
\tilde{\delta} X^I = 0~,
\label{kinst1}
\end{equation}
while the fermions just shift as
\begin{eqnarray}
\tilde{\delta} \psi^1_-
 &=& \sqrt{2 \alpha' p^+} 
    \left(
     \cos \left( \frac{m}{3} \tau \right)
       \tilde{\epsilon}^1_-
    +\sin \left( \frac{m}{3} \tau \right) \gamma^{123}
       \tilde{\epsilon}^2_+
    \right)~,
\nonumber \\
\tilde{\delta} \psi^2_+
 &=& \sqrt{2 \alpha' p^+} 
    \left(
     \cos \left( \frac{m}{3} \tau \right)
       \tilde{\epsilon}^2_+
    +\sin \left( \frac{m}{3} \tau \right) \gamma^{123}
       \tilde{\epsilon}^1_-
    \right)~,
\nonumber \\
\tilde{\delta} \psi^1_+
 &=& \sqrt{2 \alpha' p^+} 
    \left(
     \cos \left( \frac{m}{6} \tau \right)
       \tilde{\epsilon}^1_+
    +\sin \left( \frac{m}{6} \tau \right) \gamma^{123}
       \tilde{\epsilon}^2_-
    \right)~,
\nonumber \\
\tilde{\delta} \psi^2_-
 &=& \sqrt{2 \alpha' p^+} 
    \left(
     \cos \left( \frac{m}{6} \tau \right)
       \tilde{\epsilon}^2_-
    +\sin \left( \frac{m}{6} \tau \right) \gamma^{123}
       \tilde{\epsilon}^1_+
    \right)~.
\label{kinst2}
\end{eqnarray}

In this paper we study various aspects of this light-cone gauge fixed
IIA superstring theory on pp-wave geometry.  In section 2, we describe
the quantization of closed string theory: the quantization of string
coordinates and of the light-cone Hamiltonian.  In section 3, we study
the worldvolume supersymmetry algebra.  It is shown that unlike the
maximally supersymmetric case, typically of \cite{met044}, the
supersymmetry algebra has slightly abnormal structure, the reason for
which is understood by contemplating the various chiralities of
supercharges and fermionic coordinates.  In section 4, we consider
open superstring theory describing D-branes. Firstly, we give the mode
expansions of open strings taking into account various boundary
conditions.  Then we find the half-BPS D$p$-branes of $p$=2, 4, 6,
8. They should be at the origin in the transverse directions to
preserve half supersymmetries.\footnote{Similar solutions in IIB
string theory on maximally supersymmetric pp-wave have been found in
\cite{dab231,ske054,bil028,bai038}. See also \cite{kum025,ali134}.}
The directions they can be stretched are also restricted.  Among
these, we identify one D4-brane configuration with the rotating
longitudinal five brane solution in matrix model \cite{hyu090}, whose
worldvolume dynamics in low energy is governed by super Yang-Mills
theory with Kahler-Chern-Simons and Myers terms \cite{hyu219}. Another
D4-brane configuration should also have counterpart in matrix
model. We give the corresponding longitudinal five brane configuration
in matrix model.  As will be clear, due to the nature of light-cone
gauge fixing, $x^{\pm}$ satisfy the Neumann boundary condition and
thus D0-brane can not be seen.  In this sense as well as for the study
of interaction, the study of covariant GS superstring theory on the
pp-wave would be anticipated. We hope to return to this issue in the
near future.

\section{Quantization of Closed Strings}

The Type IIA string action in the light-cone gauge, (\ref{lc-action}),
is the theory of two $(4,4)$ supermultiplets.  Since it is a free
field theory, it can be solved exactly.  In this section, we will
perform quantization of the theory and obtain the light-cone
Hamiltonian which gives the exact spectrum.

The quantization begins with mode expansion of fields in the theory,
which is given in terms of solutions of the field equations.  We first
consider the bosonic sector of the theory.  The equations of motion
for the bosonic coordinates $X^I$ are read off from the action
(\ref{lc-action}) as
\begin{eqnarray}
& & \eta^{mn} \partial_m \partial_n X^i 
  - \left( \frac{m}{3} \right)^2 X^i = 0 ~,
                                    \nonumber \\
& & \eta^{mn} \partial_m \partial_n X^{i'} 
  - \left( \frac{m}{6} \right)^2 X^{i'} = 0~,
\label{beom}
\end{eqnarray}
where the fields are subject to the periodic boundary condition,
\begin{equation}
X^I (\tau, \sigma+2 \pi) = X^I (\tau, \sigma)~.
\end{equation}
The solutions of the above equations are characterized by discrete
momentum, say $n$, in the $\sigma$ direction due to the periodic
boundary condition and have the form of plane-wave.  For $X^i$, we
have two mode solutions for each $n$, which are
\begin{equation}
\phi_n (\tau, \sigma) = e^{-i \omega_n \tau - i n \sigma}~,~~~
\tilde{\phi}_n (\tau, \sigma) = e^{-i \omega_n \tau + i n \sigma}~,
\label{mode}
\end{equation}
with the wave frequency
\begin{equation}
\omega_n = {\rm sign}(n) 
      \sqrt{ \left( \frac{m}{3} \right)^2 + n^2 }~.
\end{equation}
We note that $\phi_n$ ($\tilde{\phi}_n$) becomes the left-moving
(right-moving) wave on the string worldsheet in the massless case,
that is, when we take the mass $m$ to vanish.  As for the coordinates
$X^{i'}$, we have the same mode solutions but with $m/6$ instead of
$m/3$.  In order to distinguish the modes of $X^{i'}$ from those of
$X^i$, the primed quantities will be used;
\begin{equation}
\phi'_n (\tau, \sigma) = e^{-i \omega'_n \tau - i n \sigma}~,~~~
\tilde{\phi}'_n (\tau, \sigma) 
  = e^{-i \omega'_n \tau + i n \sigma}~,~~~
\omega'_n = {\rm sign}(n) 
      \sqrt{ \left( \frac{m}{6} \right)^2 + n^2 }~.
\label{pmode}
\end{equation}

With the solutions of the equations of motion, Eq.~(\ref{beom}), the
mode expansions of the string coordinates $X^I$ are then given by
\begin{eqnarray}
X^i (\tau,\sigma) 
  &=& x^i \cos \left( \frac{m}{3} \tau \right) 
     + \alpha' p^i \frac{3}{m}
             \sin \left( \frac{m}{3} \tau \right) 
   + i \sqrt{ \frac{\alpha'}{2} } \sum_{n \neq 0}
         \frac{1}{\omega_n} 
      ( \alpha^i_n \phi_n (\tau, \sigma )
    + \tilde{\alpha}^i_n \tilde{\phi}_n (\tau, \sigma) ) ~,
                                   \nonumber \\
X^{i'} (\tau,\sigma) 
  &=& x^{i'} \cos \left( \frac{m}{6} \tau \right)
     + \alpha' p^{i'} \frac{6}{m}
              \sin \left( \frac{m}{6} \tau \right) 
   + i \sqrt{ \frac{\alpha'}{2} } \sum_{n \neq 0}
         \frac{1}{\omega'_n} 
      ( \alpha^{i'}_n \phi'_n (\tau, \sigma)
     + \tilde{\alpha}^{i'}_n \tilde{\phi}'_n (\tau, \sigma) )~,
                                   \nonumber \\
\label{bmode}
\end{eqnarray}
where $x^I$ and $p^I$ are center-of-mass variables defined in the
usual manner, coefficients for the zero-modes, and $\alpha^I_n$ and
$\tilde{\alpha}^I_n$ are the expansion coefficients for the non-zero
modes.  Here the reality of $X^I$ requires that $\alpha^{I \dagger}_n
= \alpha^I_{-n}$ and $\tilde{\alpha}^{I \dagger}_n =
\tilde{\alpha}^I_{-n}$.

We now promote the expansion coefficients in the mode expansions
(\ref{bmode}) to operators and give the commutation relations between
them leading to the quantization of bosonic fields.  The canonical
equal time commutation relations for the bosonic fields are
\begin{equation}
[ X^I (\tau, \sigma), {\cal P}^J (\tau, \sigma') ]
= i \delta^{IJ} \delta (\sigma - \sigma') ~,
\label{bcom}
\end{equation}
where ${\cal P}^J = \partial_\tau X^J / 2 \pi \alpha'$ is the
canonical conjugate momentum of $X^J$.  By using these commutators,
the mode expansions, (\ref{bmode}), lead us to have the following
commutation relations:
\begin{equation}
[ x^I, p^J ] = i \delta^{IJ} ~,~~~
[ \alpha^i_n, \alpha^j_m ] = \omega_n \delta^{ij} \delta_{n+m,0}~,~~~
[ \alpha^{i'}_n, \alpha^{j'}_m ] = 
   \omega'_n \delta^{i' j'} \delta_{n+m,0}~.
\label{bmcom}
\end{equation}

Let us next turn to the fermionic sector of the theory.  As introduced
in the last section, the fermionic fields are split into two parts
according to the $(4,4)$ supersymmetry; $(\psi^1_-, \psi^2_+)$ and
$(\psi^1_+, \psi^2_-)$.  We first consider the former case.  The
equations of motion for $\psi^1_-$ and $\psi^2_+$ are obtained as
\begin{eqnarray}
& & \partial_+ \psi^1_- + \frac{m}{3} \gamma^4 \psi^2_+ = 0 ~,
                                      \nonumber \\
& & \partial_- \psi^2_+ - \frac{m}{3} \gamma^4 \psi^1_- = 0 ~.
\label{feom}
\end{eqnarray}
The non-zero mode solutions of these equations are given by using the
modes, (\ref{mode}). For the zero mode part of the solution, we impose
a condition that, at $\tau=0$, the solution behaves just as that of
massless case.  The mode expansions for the fermionic coordinates are
then
\begin{eqnarray}
\psi^1_- (\tau, \sigma) 
 &=& c_0 \tilde{\psi}_0 \cos \left( \frac{m}{3} \tau \right) 
     - c_0 \gamma^4 \psi_0 \sin \left( \frac{m}{3} \tau \right) 
                              \nonumber \\
 & & + \sum_{n \neq 0} c_n 
      \left( 
         \tilde{\psi}_n \tilde{\phi}_n (\tau, \sigma)
         - i \frac{3}{m} (\omega_n - n) 
            \gamma^4 \psi_n \phi_n (\tau, \sigma ) 
      \right)~,
                               \nonumber \\
\psi^2_+ (\tau, \sigma) 
 &=& c_0 \psi_0 \cos \left( \frac{m}{3} \tau \right) 
     + c_0 \gamma^4 \tilde{\psi}_0 
        \sin \left( \frac{m}{3} \tau \right) 
                              \nonumber \\
 & & + \sum_{n \neq 0} c_n 
      \left( 
         \psi_n \phi_n (\tau, \sigma)
         + i \frac{3}{m} (\omega_n - n) 
            \gamma^4 \tilde{\psi}_n \tilde{\phi}_n (\tau, \sigma ) 
      \right)~,
\label{fmode}
\end{eqnarray}
where $\gamma^{1234} \psi_n = \psi_n$ and $\gamma^{1234}
\tilde{\psi}_n = -\tilde{\psi}_n$ for all $n$, and $c_n$ are the
normalization constants to be fixed in the process of quantization.
As in the case of bosonic sector, the expansion coefficients $\psi_n$
and $\tilde{\psi}_n$ are promoted to operators.  By using the
canonical equal time anti-commutation relations,
\begin{equation}
\{ \psi^A_\pm (\tau, \sigma), \psi^B_\pm (\tau, \sigma') \}
= 2 \pi \alpha' \delta^{AB} \delta (\sigma - \sigma') ~,
\label{fcom}
\end{equation}
and fixing the normalization constants as
\[
c_0 = \sqrt{\alpha'}~,~~~
c_n = \frac{\sqrt{\alpha'}}{\sqrt{1 + 
          \left( \frac{3}{m} \right)^2 (\omega_n - n)^2 } }~,
\]
the following anti-commutation relations between mode operators are
obtained.
\begin{equation}
\{ \psi_n , \psi_m \} = \delta_{n+m,0} ~,~~~
\{ \tilde{\psi}_n, \tilde{\psi}_m \} = \delta_{n+m,0}~.
\label{fmcom}
\end{equation}

The quantization of fermionic coordinates $(\psi^1_+, \psi^2_-)$ in
the other $(4,4)$ supermultiplet proceeds along the same way with that
of above case.  The equations of motion for $\psi^1_+$ and $\psi^2_-$
are respectively
\begin{eqnarray}
& & \partial_+ \psi^1_+ - \frac{m}{6} \gamma^4 \psi^2_- = 0 ~,
                                      \nonumber \\
& & \partial_- \psi^2_- + \frac{m}{6} \gamma^4 \psi^1_+ = 0 ~.
\label{fpeom}
\end{eqnarray}
In the present case, the basic solutions are given in terms of the
primed ones, (\ref{pmode}), and the mode expansions for the fields are
\begin{eqnarray}
\psi^1_+ (\tau, \sigma) 
 &=& c'_0 \tilde{\psi}'_0 \cos \left( \frac{m}{6} \tau \right) 
     + c'_0 \gamma^4 \psi'_0 \sin \left( \frac{m}{6} \tau \right) 
                              \nonumber \\
 & & + \sum_{n \neq 0} c'_n 
      \left( 
         \tilde{\psi}'_n \tilde{\phi}'_n (\tau, \sigma)
         + i \frac{6}{m} (\omega'_n - n) 
            \gamma^4 \psi'_n \phi'_n (\tau, \sigma ) 
      \right)~,
                               \nonumber \\
\psi^2_- (\tau, \sigma) 
 &=& c'_0 \psi'_0 \cos \left( \frac{m}{6} \tau \right) 
     - c'_0 \gamma^4 \tilde{\psi}'_0 
        \sin \left( \frac{m}{6} \tau \right) 
                              \nonumber \\
 & & + \sum_{n \neq 0} c'_n 
      \left( 
         \psi'_n \phi'_n (\tau, \sigma)
         - i \frac{6}{m} (\omega'_n - n) 
            \gamma^4 \tilde{\psi}'_n \tilde{\phi}'_n (\tau, \sigma ) 
      \right)~,
\label{fpmode}
\end{eqnarray}
where $\gamma^{1234} \psi'_n = -\psi'_n$ and $\gamma^{1234}
\tilde{\psi}'_n = \tilde{\psi}'_n$.  Then the equal time
anti-commutation relations, (\ref{fcom}), and the following fixing of
the normalization constants
\[
c'_0 = \sqrt{\alpha'}~,~~~
c'_n = \frac{\sqrt{\alpha'}}{\sqrt{1 + 
          \left( \frac{6}{m} \right)^2 (\omega'_n - n)^2 } }~,
\]
lead us to the anti-commutators between mode operators.
\begin{equation}
\{ \psi'_n , \psi'_m \} = \delta_{n+m,0} ~,~~~
\{ \tilde{\psi}'_n, \tilde{\psi}'_m \} = \delta_{n+m,0}~.
\label{fpmcom}
\end{equation}

Having the quantized $(4,4)$ supermultiplets which are free, we now
consider the light-cone Hamiltonian of the theory which has the
quadratic form in term of the string coordinates or mode operators and
is thus exact.  The light-cone Hamiltonian is written as\footnote{
From now on, we set $2 \pi \alpha' = 1$ for notational convenience.}
\begin{equation}
H_{LC} 
 = \frac{1}{2 \pi} \int^{2\pi}_0 d \sigma P^-
 = \frac{1}{p^+} \int^{2\pi}_0 d \sigma {\cal H} ~,
\label{lc-ham}
\end{equation}
where ${\cal H}$ is the Hamiltonian density of the light-cone gauge
fixed string action, Eq.~(\ref{lc-action}), and is obtained as
\begin{eqnarray}
{\cal H} 
&=&  \frac{1}{2} ({\cal P}^I)^2 
    + \frac{1}{2} (\partial_\sigma X^I)^2
    +\frac{1}{2} \left( \frac{m}{3} \right)^2 (X^i)^2
    +\frac{1}{2} \left( \frac{m}{6} \right)^2 (X^{i'})^2
  \nonumber \\
 & & - \frac{i}{2} \psi^1_- \partial_\sigma \psi^1_- 
     + \frac{i}{2} \psi^2_+ \partial_\sigma \psi^2_+
     + i \frac{m}{3} \psi^2_+ \gamma^4 \psi^1_-
  \nonumber \\
 & & - \frac{i}{2} \psi^1_+ \partial_\sigma \psi^1_+
     + \frac{i}{2} \psi^2_- \partial_\sigma \psi^2_-
     - i \frac{m}{6} \psi^2_- \gamma^4 \psi^1_+ ~.
\label{can-h}
\end{eqnarray} 
By plugging the mode expansions for the fields, Eqs. (\ref{bmode}),
(\ref{fmode}), and (\ref{fpmode}), into Eq. (\ref{lc-ham}), we see
that the light-cone Hamiltonian becomes
\begin{equation}
H_{LC} = E_0 + E + \tilde{E}~,
\label{lc-h}
\end{equation}
where $E_0$ is the zero mode contribution and $E$ ($\tilde{E}$) the
contribution of the non-zero modes of the type $\alpha^I_n$, $\psi_n$,
and $\psi'_n$ ( $\tilde{\alpha}^I_n$, $\tilde{\psi}_n$,
$\tilde{\psi}'_n$).  Each of the contributions is expressed as
follows:
\begin{eqnarray}
E_0 &=& \frac{\pi}{p^+} 
  \left( \left( \frac{p^I}{2\pi} \right)^2 
       + \left( \frac{m}{3} \right)^2 (x^i)^2
       + \left( \frac{m}{6} \right)^2 (x^{i'})^2
       - \frac{i}{\pi} \frac{m}{3} \tilde{\psi}_0 \gamma^4 \psi_0
       + \frac{i}{\pi} \frac{m}{6} \tilde{\psi}'_0 \gamma^4 \psi'_0
  \right)~,
  \nonumber \\
E &=& \frac{1}{2 p^+} \sum_{n \neq 0} 
  ( \alpha^I_{-n} \alpha^I_n  
   + \omega_n \psi_{-n} \psi_n
   + \omega'_n \psi'_{-n} \psi'_n
  )~,
  \nonumber \\
\tilde{E}
  &=& \frac{1}{2 p^+} \sum_{n \neq 0} 
  ( \tilde{\alpha}^I_{-n} \tilde{\alpha}^I_n  
   + \omega_n \tilde{\psi}_{-n} \tilde{\psi}_n
   + \omega'_n \tilde{\psi}'_{-n} \tilde{\psi}'_n
  )~.
\end{eqnarray}

In the quantized version, the modes in the expression of Hamiltonian
become operators with the commutation relations, (\ref{bmcom}),
(\ref{fmcom}), and (\ref{fpmcom}), and should be properly normal
ordered.  For the non-zero mode or the string oscillator
contributions, that is, $E$ and $\tilde{E}$, we place operator with
negative mode number to the left of operator with positive mode number
as in the flat case.  This is natural since $E$ and $\tilde{E}$
become those of the string in the flat background when the mass $m$ is
taken to vanish.  The normal ordered expressions of them are then
given by
\begin{eqnarray}
E &=& \frac{1}{p^+} \sum^\infty_{n =1} 
  ( \alpha^I_{-n} \alpha^I_n  
   + \omega_n \psi_{-n} \psi_n
   + \omega'_n \psi'_{-n} \psi'_n
  )~,
  \nonumber \\
\tilde{E}
  &=& \frac{1}{p^+} \sum^\infty_{n =1} 
  ( \tilde{\alpha}^I_{-n} \tilde{\alpha}^I_n  
   + \omega_n \tilde{\psi}_{-n} \tilde{\psi}_n
   + \omega'_n \tilde{\psi}'_{-n} \tilde{\psi}'_n
  )~.
\label{nzero-h}
\end{eqnarray}
Here we note that there is no zero-point energy because bosonic
contributions are exactly canceled by those of fermions.

The zero mode contribution is the Hamiltonian for the simple harmonic
oscillators and massive fermions.  For the bosonic part, we introduce
the usual creation and annihilation operators as
\begin{eqnarray}
& & a^{i \dagger} = \sqrt{ \frac{3 \pi}{m} }  
      \left( \frac{p^i}{2 \pi} + i \frac{m}{3} x^i \right)~,
 ~~~a^i = \sqrt{ \frac{3 \pi}{m} }
      \left( \frac{p^i}{2 \pi} - i \frac{m}{3} x^i \right)~,
  \nonumber \\
& & a^{i' \dagger} = \sqrt{ \frac{6 \pi}{m} }
    \left( \frac{p^{i'}}{2 \pi} + i \frac{m}{6} x^{i'} \right)~,
 ~~~a^{i'} = \sqrt{ \frac{6 \pi}{m} }
    \left( \frac{p^{i'}}{2 \pi} - i \frac{m}{6} x^{i'} \right)~,
\label{bz}
\end{eqnarray}
whose commutation relations are read as, from Eq.~(\ref{bmcom}),
\begin{equation}
[ a^I, a^{J \dagger} ] = \delta^{IJ}~.
\end{equation}
As for the fermionic creation and annihilation operators, we take the
following combination of modes, the reason for which will be explained
below.
\begin{eqnarray}
& & \chi^\dagger = \frac{1}{\sqrt{2}} 
             ( \psi_0 - i \gamma^4 \tilde{\psi}_0 )~,
 ~~~\chi = \frac{1}{\sqrt{2}} 
             ( \psi_0 + i \gamma^4 \tilde{\psi}_0 )~,
  \nonumber \\
& & \chi'^\dagger = \frac{1}{\sqrt{2}} 
             ( \psi'_0 + i \gamma^4 \tilde{\psi}'_0 )~,
 ~~~\chi' = \frac{1}{\sqrt{2}} 
             ( \psi'_0 - i \gamma^4 \tilde{\psi}'_0 )~,
\label{fz}
\end{eqnarray}
where $\gamma^{12349} \chi = - \chi$ and $\gamma^{12349} \chi' = 
\chi'$.  From Eqs.~(\ref{fmcom}) and (\ref{fpmcom}), the
anti-commutation relations between these operators become
\begin{equation}
\{ \chi, \chi^\dagger \} = 1 ~, ~~~
\{ \chi', \chi'^\dagger \} = 1 ~,
\end{equation}
where, since each of the fermionic coordinates has four independent
components although we have used the 16 component notation, the right
hand sides should be understood as $4 \times 4$ unit matrix.  In terms
of the operators introduced above, Eqs.~(\ref{bz}) and (\ref{fz}), the
normal ordered zero mode contribution to the light-cone Hamiltonian is
then given by
\begin{equation}
E_0 = \frac{m}{6 p^+} 
     ( 2 a^{i \dagger} a^i + a^{i' \dagger} a^{i'}
       + 2 \chi^\dagger \chi + \chi'^\dagger \chi' )~.
\label{zero-h}
\end{equation}
Here we see that $E_0$ has vanishing zero-point energy as in the case
of string oscillator contributions.  The zero-point energy of the
bosonic part is evaluated as
\[
\frac{1}{2} \left( 4 \times \frac{m}{3} + 4 \times \frac{m}{6} \right)
= \frac{m}{6} \times 6 ~,
\]
which is exactly canceled by that from the fermionic part.

The normal ordered expressions Eqs.~(\ref{nzero-h}) and (\ref{zero-h})
now constitute the quantum light-cone Hamiltonian, which implicitly
defines the vacuum $|0 \rangle$ of the quantized theory as a state
annihilated by string oscillation operators with positive mode number,
that is $n \ge 1$, and zero mode operators $a^I$, $\chi$, and $\chi'$
defined in Eqs.~(\ref{bz}) and (\ref{fz}).  Actually, the vacuum
defined in this paper, especially the vacuum state in the zero mode
sector, is not unique but one of the possible Clifford vacua, since
our theory is massive and there can be various definitions for the
creation and annihilation operators.  This is also the case for the
IIB superstring in pp-wave background and has been discussed in
\cite{met109}.  However, considering the regularity of states at $\tau
\rightarrow i \infty$ that has been pointed out in \cite{rus179}, our
definition is a natural one. This is because the expansion
coefficients corresponding to mode solutions diverging at $\tau
\rightarrow i \infty$ in Eqs.~(\ref{bmode}), (\ref{fmode}), and
(\ref{fpmode}) must annihilate the vacuum in order to ensure the
regularity of physical states constructed out from the vacuum at such
Euclidean time region.

The string states are then obtained by acting creation operators on
the vacuum $| 0 \rangle$.  However, the physical states are not all
possible such states but those in the subspace of states which are
constrained by the Virasoro constraint imposing the invariance under
the translation in $\sigma$ direction.  In the light-cone gauge, 
the Virasoro constraint is given by
\begin{equation}
\int^{2\pi}_0 d \sigma 
 \left( - \frac{1}{2\pi} p^+ \partial_\sigma X^-
   + {\cal P}^I \partial_\sigma X^I
   + \frac{i}{2} \psi^A_+ \partial_\sigma \psi^A_+
   + \frac{i}{2} \psi^A_- \partial_\sigma \psi^A_-
 \right) = 0 ~.
\label{vira}
\end{equation}
The integration of the first integrand vanishes trivially since $p^+$
is constant, and the remaining parts give us the following constraint.
\begin{equation}
N = \tilde{N} ~,
\end{equation}
where $N$ and $\tilde{N}$ are defined as
\begin{eqnarray}
N &=& \sum^\infty_{n=1} n 
   \left( \frac{1}{\omega_n} \alpha^i_{-n} \alpha^i_n
         +\frac{1}{\omega'_n} \alpha^{i'}_{-n} \alpha^{i'}_n
         + \psi_{-n} \psi_n + \psi'_{-n} \psi'_n
   \right) ~,
  \nonumber \\
\tilde{N}
  &=&\sum^\infty_{n=1} n
   \left( \frac{1}{\omega_n} \tilde{\alpha}^i_{-n} 
                             \tilde{\alpha}^i_n
         +\frac{1}{\omega'_n} \tilde{\alpha}^{i'}_{-n} 
                              \tilde{\alpha}^{i'}_n
         + \tilde{\psi}_{-n} \tilde{\psi}_n 
         + \tilde{\psi}'_{-n} \tilde{\psi}'_n
   \right) ~.
\end{eqnarray}
Here we normal ordered the expressions and see that the normal
ordering constants have canceled between bosonic and fermionic
contributions.

\section{Supersymmetry Algebra}

The light-cone gauge fixed action (\ref{lc-action}) is invariant under
the dynamical and the kinematical supersymmetry transformation.  The
dynamical (kinematical) supersymmetry parameter has eight (sixteen)
independent components hence leading to eight (sixteen)
supersymmetries.  In this section, we obtain the supercharges
corresponding to these supersymmetries and compute the supersymmetry
algebra between them.

The charges for the supersymmetry transformations, (\ref{dynst1}),
(\ref{dynst2}), (\ref{kinst1}) and (\ref{kinst2}), are derived through
the standard Noether procedure.  For the kinematical supersymmetry, we
have four types of supercharge, $\widetilde{Q}^A_\pm$, corresponding
to the transformation parameters, $\tilde{\epsilon}^A_\pm$, which are
derived as
\begin{eqnarray}
\widetilde{Q}^1_-
  &=& \sqrt{ \frac{p^+}{\pi} } \int^{2 \pi}_0 d \sigma
      \left( \cos \left( \frac{m}{3} \tau \right)
               \psi^1_-
            -\sin \left( \frac{m}{3} \tau \right) \gamma^{123}
               \psi^2_+
      \right) ~,
\nonumber \\
\widetilde{Q}^2_+
  &=& \sqrt{ \frac{p^+}{\pi} } \int^{2 \pi}_0 d \sigma
      \left( \cos \left( \frac{m}{3} \tau \right)
               \psi^2_+
            -\sin \left( \frac{m}{3} \tau \right) \gamma^{123}
               \psi^1_-
      \right) ~,
\nonumber \\
\widetilde{Q}^1_+
  &=& \sqrt{ \frac{p^+}{\pi} } \int^{2 \pi}_0 d \sigma
      \left( \cos \left( \frac{m}{6} \tau \right)
               \psi^1_+
            -\sin \left( \frac{m}{6} \tau \right) \gamma^{123}
               \psi^2_-
      \right) ~,
\nonumber \\
\widetilde{Q}^2_-
  &=& \sqrt{ \frac{p^+}{\pi} } \int^{2 \pi}_0 d \sigma
      \left( \cos \left( \frac{m}{6} \tau \right)
               \psi^2_-
            -\sin \left( \frac{m}{6} \tau \right) \gamma^{123}
               \psi^1_+
      \right) ~.
\label{kcharge}
\end{eqnarray}
It is now convenient to introduce combinations of these supercharges
as
\begin{equation}
\widetilde{Q}^\pm = \widetilde{Q}^1_\pm + \widetilde{Q}^2_\mp
\end{equation}
such that the eigenvalue of $\gamma^{12349}$ or of $\gamma^{5678}$
becomes manifest.  The kinematical supersymmetry transformations rules
of string coordinates, (\ref{kinst1}) and (\ref{kinst2}), are then
obtained from
\begin{equation}
\tilde{\delta} \varphi 
 = [ \tilde{\epsilon}^+ \widetilde{Q}^+
    +\tilde{\epsilon}^- \widetilde{Q}^- , \varphi ] ~,
\end{equation}
where $\tilde{\epsilon}^\pm = \tilde{\epsilon}^1_\pm +
\tilde{\epsilon}^2_\mp$.  It is easily seen that one $(4,4)$
supermultiplet $(X^i, \psi^1_-, \psi^2_+)$ is only affected by
$\widetilde{Q}^-$ and the other supermultiplet $(X^{i'}, \psi^1_+,
\psi^2_-)$ only by $\widetilde{Q}^+$.

As for the dynamical supersymmetry transformation with parameters,
$\epsilon^1_+$ and $\epsilon^2_-$, we have obtained the following two
types of supercharge.
\begin{eqnarray}
Q^1_+ 
 &=& \sqrt{ \frac{ 2\pi}{p^+} }
    \int^{2 \pi}_0 d \sigma
     \left( \partial_- X^i \gamma^i \psi^1_- 
           + \partial_- X^{i'} \gamma^{i'} \psi^1_+
           + \frac{m}{3} X^i \gamma^i \gamma^4 \psi^2_+
           - \frac{m}{6} X^{i'} \gamma^{i'} \gamma^4 \psi^2_-
     \right) ~,
\nonumber \\
Q^2_- 
  &=& \sqrt{ \frac{ 2\pi}{p^+} }
    \int^{2 \pi}_0 d \sigma
     \left( \partial_+ X^i \gamma^i \psi^2_+ 
           + \partial_+ X^{i'} \gamma^{i'} \psi^2_-
           - \frac{m}{3} X^i \gamma^i \gamma^4 \psi^1_-
           + \frac{m}{6} X^{i'} \gamma^{i'} \gamma^4 \psi^1_+
     \right) ~.
\label{dcharge}
\end{eqnarray}
Since two supercharges have different transverse $SO(8)$ chirality,
that is, $Q^1_+$ is in ${\bf 8_c}$ of $SO(8)$ and $Q^2_-$ in ${\bf
  8_s}$, and each of them has four independent spinor components, they
naturally represent the dynamical $(4,4)$ supersymmetry as it should
be.  If we now combine them so as to make the eigenvalue of
$\gamma^{12349}$ manifest as done in the kinematical case
\begin{equation}
Q^- = Q^1_+ + Q^2_- ~,
\end{equation}
we can check that the dynamical supersymmetry transformation rules,
(\ref{dynst1}) and (\ref{dynst2}), follow through the relation
\begin{equation}
\delta \varphi = [ \epsilon^- Q^- , \varphi ] ~,
\end{equation}
where $\epsilon^- = \epsilon^1_+ + \epsilon^2_-$.

Having derived the supercharges, (\ref{kcharge}) and (\ref{dcharge}),
let us turn to the computation of the supersymmetry algebra between
them.  The basic rules for it are the commutation relations of
Eqs.~(\ref{bcom}) and (\ref{fcom}), and the $SO(8)$ Fierz identity
being obtained from the usual $SO(9)$ identity.

The first algebra we consider is that between kinematical
supercharges, which is simply computed as
\begin{equation}
\{ \widetilde{Q}^\pm_\alpha, \widetilde{Q}^\pm_\beta \} 
= 2 p^+ (h_\pm)_{\alpha \beta} ~,
\end{equation}
where $\alpha, \beta$ are spinor indices and $h_\pm$ are projection
operators onto spinor states of positive and negative eigenvalue of 
$\gamma^{12349}$ respectively,
\begin{equation}
h_\pm = \frac{1}{2} ( 1 \pm \gamma^{12349} ) ~.
\end{equation}

For the supersymmetry algebra between kinematical and dynamical 
supercharges, we get
\begin{eqnarray}
\{ \widetilde{Q}^+_\alpha, Q^-_\beta \}
 &=& \frac{1}{\sqrt{2}} (h_+ \gamma^{i'})_{\alpha \beta} P^{i'}
    -\frac{1}{\sqrt{2}}
      \frac{\mu}{6} (h_+ \gamma^{123} \gamma^{i'})_{\alpha\beta}
       J^{+i'} ~,
  \nonumber \\
\{ \widetilde{Q}^-_\alpha, Q^-_\beta \}
 &=& \frac{1}{\sqrt{2}} (h_- \gamma^i)_{\alpha\beta} P^i
    -\frac{1}{\sqrt{2}}
       \frac{\mu}{3} ( h_- \gamma^{123} \gamma^i)_{\alpha\beta}
       J^{+i} ~,
\end{eqnarray}
where we have defined the following quantities.
\begin{eqnarray}
P^i &=& \int d \sigma 
  \left( \cos \left(\frac{m}{3} \tau \right) {\cal P}^i 
        + \frac{m}{3} \sin \left( \frac{m}{3} \tau \right) X^i
  \right) ~,
  \nonumber \\
P^{i'} &=& \int d \sigma 
  \left( \cos \left(\frac{m}{6} \tau \right) {\cal P}^{i'} 
        + \frac{m}{6} \sin \left( \frac{m}{6} \tau \right) X^{i'}
  \right) ~,
  \nonumber \\
J^{+i} &=& \int d \sigma
  \left( \frac{3}{\mu}
         \sin \left(\frac{m}{3} \tau \right) {\cal P}^i
        - \frac{1}{2 \pi} \cos \left(\frac{m}{3} \tau \right) 
          p^+ X^i
  \right) ~,
  \nonumber \\
J^{+i'} &=& \int d \sigma
  \left( \frac{6}{\mu}
         \sin \left(\frac{m}{6} \tau \right) {\cal P}^{i'} 
        - \frac{1}{2 \pi} \cos \left(\frac{m}{6} \tau \right) 
          p^+ X^{i'}
  \right) ~.
\end{eqnarray}
$P^I$ and $J^{+I}$ are the nothing but the kinematical generators for
the translation in the transverse space and the rotation in the $(x^-,
x^I)$ plane respectively, which have been extensively discussed in
\cite{met044}.

Finally, the dynamical supersymmetry algebra is computed as
\begin{equation}
\{ Q^-_\alpha , Q^-_\beta \} 
  = 4 \pi (h_-)_{\alpha \beta} H_{LC}
     - \frac{\mu}{3}
        (h_- \gamma^{ij} \gamma^{123})_{\alpha \beta} J^{ij}
     + \frac{\mu}{6} (h_- \gamma^{i'j'} \gamma^{123})_{\alpha \beta} 
         \widehat{J}^{i'j'} ~,
\label{d-alg}
\end{equation}
where $J^{ij}$ is the rotation generator in three dimensional space
spanned by $x^{1,2,3}$ (and thus the indices $i,j$ take values only
among $1,2,3$), and, interestingly enough, $\widehat{J}^{i'j'}$ is one
part of the rotation generator $J^{i'j'}$ in the space spanned by
$x^{5,6,7,8}$.  The structure of this algebra reflects the fact that
the symmetry group $SO(8)$ of transverse space is broken to $SO(3)
\times SO(4)$ due to the presence of the Ramond-Ramond four form and
two form field strengths, $F_{+123}$ and $F_{+4}$. The explicit
expressions of $J^{ij}$ for $SO(3)$ and $\widehat{J}^{i'j'}$ for
$SO(4)$ are given by
\begin{eqnarray}
J^{ij} &=& \int d \sigma
  \left(
  X^i {\cal P}^j - X^j {\cal P}^i 
  - \frac{i}{4} \psi^1_- \gamma^{ij} \psi^1_-
  - \frac{i}{4} \psi^2_+ \gamma^{ij} \psi^2_+
 - \frac{i}{4} \psi^1_+ \gamma^{ij} \psi^1_+
  - \frac{i}{4} \psi^2_- \gamma^{ij} \psi^2_-
  \right) ~,
  \nonumber \\
\widehat{J}^{i'j'} &=& \int d \sigma
  \left(
   X^{i'} {\cal P}^{j'} - X^{j'} {\cal P}^{i'} 
  - \frac{i}{4} \psi^1_- \gamma^{i'j'} \psi^1_-
  - \frac{i}{4} \psi^2_+ \gamma^{i'j'} \psi^2_+
  \right) ~.
\end{eqnarray}

In order to see what makes special thing in $x^{5,6,7,8}$ directions,
let us first consider the property of $\widehat{J}^{i'j'}$.  In the
above expression for $\widehat{J}^{i'j'}$, there is no dependence on
the fermionic coordinates $\psi^1_+$ and $\psi^2_-$, which states that
those coordinates do not rotate under the action of
$\widehat{J}^{i'j'}$.  The full form of the rotation generator
$J^{i'j'}$ which rotates $\psi^1_+$ and $\psi^2_-$ as well as other
string coordinates is simply given by
\begin{equation}
J^{i'j'} =
\widehat{J}^{i'j'} 
- \frac{i}{4} \int d \sigma ( \psi^1_+ \gamma^{i'j'}
 \psi^1_+ + \psi^2_- \gamma^{i'j'} \psi^2_- ) ~.
\end{equation}
Since the theory we study is non-interacting, $\widehat{J}^{i'j'}$ is
really a rotation generator but without affecting $\psi^1_+$ and
$\psi^2_-$.  The reason for why only one sub-generator rather than
full of the rotation generator $J^{i'j'}$ appears in the algebra
(\ref{d-alg}) may be understandable by looking at the chirality
structure of the dynamical supercharges and the fermionic coordinates,
which is listed in table \ref{t1}.  The right hand side of
(\ref{d-alg}) is basically the sum of field bilinears and should
respect the chirality structure of $\{ Q^-, Q^- \}$.  The $SO(4)$
rotation is sensitive to the chirality of fermionic coordinates
measured by $\gamma^{5678}$.  From the table \ref{t1}, one may
recognize that bilinears made of $\psi^1_+$ or $\psi^2_-$ cannot
contribute to the structure associated with the $SO(4)$ rotation.  On
the other hand, for the $SO(3)$ rotation where eigenvalues of
$\gamma^{1234}$ enter the story, they can contribute.  This at first
glance abnormal situation is basically because the supersymmetry is
not maximal but $(4,4)$.  If the theory had maximal dynamical
supersymmetry, we would obtain the algebra containing full of the
rotation generators as was the case of IIB superstring in pp-wave
background \cite{met044}.

\begin{table}
\begin{center}
\begin{tabular}{cccc}
\hline
   & $\gamma^9$ & $\gamma^{1234}$ & $\gamma^{5678}$ \\
\hline
 $Q^1_+$    & $-$ & $+$ & $-$ \\
 $Q^2_-$    & $+$ & $-$ & $-$ \\
\hline
 $\psi^1_-$ & $+$ & $-$ & $-$ \\
 $\psi^2_+$ & $-$ & $+$ & $-$ \\
\hline
 $\psi^1_+$ & $+$ & $+$ & $+$ \\
 $\psi^2_-$ & $-$ & $-$ & $+$ \\
\hline
\end{tabular}
\end{center}
\caption{Eigenvalues or chiralities of dynamical supercharges and 
fermionic coordinates for $\gamma^9$, $\gamma^{1234}$ and 
$\gamma^{5678}$.}
\label{t1}
\end{table}

\section{Supersymmetric D-branes}

In the light-cone framework, we now investigate what kinds of D-brane
are possible in the theory and how amount of supersymmetry is
preserved by studying the open superstring ending on D-branes.
Considering the tensor rank of Ramond-Ramond gauge field in the IIA
string theory, one expects D$p$-branes with even $p$ ranging from 0 to
8.  As is well known, however, the light-cone formulation does not
give definite decision on the presence of D0-brane, since $X^-$ always
satisfies the Neumann boundary condition due to the Virasoro
constraint, Eq.~(\ref{vira}).  Thus we are led to concentrate on the
case of $p \ge 2$.

\subsection{Quantization of Open Strings and D-branes}

The equations of motion for the open superstring are the same with
those of the closed string, that is, Eqs.~(\ref{beom}), (\ref{feom})
and (\ref{fpeom}).  In addition to them, the open string should
satisfies appropriate open string boundary conditions.  We let the end
points of open string are at $\sigma = 0 $ and $\sigma = \pi$.  Since
we will consider single brane in this paper, two end points should
satisfy the same boundary condition.  Of course, an open string may
have different boundary conditions at two ends for the case of
multiple D-branes.

We impose the Neumann boundary conditions on the longitudinal
directions of D-brane and the Dirichlet boundary conditions on the
transverse directions which are respectively
\begin{equation}
\partial_\sigma X^{I_N} |_{\sigma = 0, \pi} = 0 ~,~~~
\partial_\tau X^{I_D} |_{\sigma = 0, \pi} = 0 ~,
\label{bbc}
\end{equation}
where the index $I_N$ ($I_D$) represents the Neumann (Dirichlet)
direction.  The mode expansions may be obtained from that of closed
string, (\ref{bmode}) by imposing these boundary conditions, which are
given by
\begin{eqnarray}
X^{i_N} &=& x^{i_N} \cos \left( \frac{m}{3} \tau \right) 
        + 2 \alpha' p^{i_N} \frac{3}{m}
             \sin \left( \frac{m}{3} \tau \right) 
   + \sqrt{ 2 \alpha' } i \sum_{n \neq 0}
         \frac{1}{\omega_n} 
       \alpha^{i_N}_n e^{-i \omega_n \tau} \cos (n \sigma) ~,
                                      \nonumber \\
X^{i'_N} &=& x^{i'_N} \cos \left( \frac{m}{6} \tau \right) 
        + 2 \alpha' p^{i'_N} \frac{6}{m}
             \sin \left( \frac{m}{6} \tau \right) 
   + \sqrt{ 2 \alpha' } i \sum_{n \neq 0}
         \frac{1}{\omega'_n} 
       \alpha^{i'_N}_n e^{-i \omega'_n \tau} \cos (n \sigma) ~,
                                      \nonumber \\
X^{i_D} &=& \sqrt{ 2 \alpha' } \sum_{n \neq 0}
         \frac{1}{\omega_n} 
       \alpha^{i_D}_n e^{-i \omega_n \tau} \sin (n \sigma) ~,
                                      \nonumber \\
X^{i'_D} &=& \sqrt{ 2 \alpha' } \sum_{n \neq 0}
         \frac{1}{\omega'_n} 
       \alpha^{i'_D}_n e^{-i \omega'_n \tau} \sin (n \sigma) ~,
\label{obmode}
\end{eqnarray}
where $n \in {\bf Z}$, $i_N$ ($i_D$) is the index for Neumann
(Dirichlet) direction among $1,2,3,4$ and $i'_N$ ($i'_D$) among
$5,6,7,8$.  Actually, we have obtained the mode expansions also for
the case where different boundary conditions are imposed at two end
points from the equations of motion (\ref{beom}) and the boundary
conditions (\ref{bbc}).  For the Neumann boundary condition at
$\sigma=0$ and the Dirichlet one at $\sigma = \pi$, they are
\begin{equation}                   
X^i = \sqrt{ 2 \alpha' } i \sum_r
         \frac{1}{\omega_r} 
       \alpha^i_r e^{-i \omega_r \tau} \cos (r \sigma) ~,~~~
X^{i'} = \sqrt{ 2 \alpha' } i \sum_r
         \frac{1}{\omega'_r} 
       \alpha^{i'}_r e^{-i \omega'_r \tau} \cos (r \sigma) ~,
\label{obmode1}
\end{equation}
and, for the case where the boundary conditions are imposed inversely,
we get
\begin{equation}
X^i = \sqrt{ 2 \alpha' } \sum_r
         \frac{1}{\omega_r} 
       \alpha^i_r e^{-i \omega_r \tau} \sin (r \sigma)~,~~~
X^{i'} = \sqrt{ 2 \alpha' } \sum_r
         \frac{1}{\omega'_r} 
       \alpha^{i'}_r e^{-i \omega'_r \tau} \sin (r \sigma) ~,
\label{obmode2}
\end{equation}
where $r \in {\bf Z} + \frac{1}{2}$.  Now it is clear from the above
mode expansions that Dirichlet brane is centered at the origin of the
pp-wave background.  We note that the same observation was given also
for the IIB case in the light-cone framework \cite{dab231}.

For the fermionic coordinates, the boundary condition is given by
\begin{equation}
\psi^1_\pm \Big|_{\sigma=0,\pi} 
 = \Omega \psi^2_\mp \Big|_{\sigma=0,\pi} ~,
\label{fbc}
\end{equation}
where, as noted in \cite{dab231}, $\Omega$ is the product of gamma
matrices with indices of Dirichlet directions.  From the mode
expansions, (\ref{fmode}) and (\ref{fpmode}), we see that the
fermionic boundary condition gives the following relations between
expansion coefficients that lead us to the mode expansion for open
string fermionic coordinates: for all $n$,
\begin{equation}
\tilde{\psi}_n = \Omega \psi_n ~,~~~
\tilde{\psi}'_n = \Omega \psi'_n ~.
\label{fmbc}
\end{equation}

Precise knowledge about $\Omega$ gives possible D-brane configurations
in the IIA pp-wave background.  If we evaluate the equations of motion
for fermions, (\ref{feom}) and (\ref{fpeom}), at boundaries, we have
the following condition for $\Omega$:
\begin{equation}
\gamma^4 \Omega \gamma^4 \Omega = -1 ~.
\label{oc}
\end{equation}
Since $\Omega$ relates fermionic coordinates with different $SO(8)$
chiralities and eigenvalues for $\gamma^{1234}$, it should
anti-commute with $\gamma^9$ and $\gamma^{1234}$ as follows.
\begin{equation}
\{ \Omega, \gamma^9 \} = 0 ~, ~~~
\{ \Omega, \gamma^{1234} \} = 0 ~.
\label{og}
\end{equation}

Now Eqs.~(\ref{oc}) and (\ref{og}) let us know the structure of
$\Omega$: First of all, $\Omega$ should be a product of odd number of
gamma matrices.  Secondly, among the gamma matrices, even number of
them should be indexed in $1,2,3,4$ and hence even number of them in
$5,6,7,8$.  Finally, since the gamma matrices are symmetric, we have
$\Omega^T \Omega = 1$.  With these informations, one can solve
Eq.~(\ref{oc}) rather easily and know about possible single D-brane
configurations, which is listed in table \ref{t2}.

\begin{table}
\begin{center}
\begin{tabular}{c|c}
\hline
 $N_D$ & $\Omega$ \\
\hline
 1 & $\gamma^i$ \\
 3 &  $\gamma^{ij}\gamma^4$, $\gamma^{ i' j'} \gamma^4$ \\
 5 &  $\gamma^{123} \gamma^{i'j'}$, $\gamma^i \gamma^{5678}$ \\
 7 &  $\gamma^{ij} \gamma^{45678}$ \\
\hline
\end{tabular}
\end{center}
\caption{List of $\Omega$ satisfying the condition, (\ref{oc}).
$N_D$ is the number of Dirichlet directions and we temporarily
restrict the indices $i,j$ to take values in $1,2,3$. }
\label{t2}
\end{table}

We now describe the light-cone Hamiltonian and its quantization for
the open superstring describing single D-brane.  Since the story will
be almost the same with that for the closed string case, the
description will be rather brief.  The light-cone Hamiltonian for the
open string is given by
\begin{equation}
H_{LC} = \frac{1}{\pi} \int^\pi_0 d \sigma P^-
       = \frac{2}{p^+} \int^\pi_0 d \sigma {\cal H} ~.
\end{equation}
where ${\cal H}$ is given in Eq.~(\ref{can-h}).  By plugging the mode
expansions of Eq.~(\ref{obmode}) for bosonic string coordinates and
those of Eqs.~(\ref{fmode}) and (\ref{fpmode}) with the condition
(\ref{fmbc}) for fermionic coordinates into the Hamiltonian, we get
\begin{equation}
H_{LC} = E_0 + E ~,
\end{equation}
where $E_0$ and $E$ are the zero mode and non-zero mode contributions
respectively.  Firstly, the normal ordered expression for $E$ is read
as
\begin{equation}
E = \frac{2}{p^+} \sum^\infty_{n =1} 
  ( \alpha^I_{-n} \alpha^I_n  
   + \omega_n \psi_{-n} \psi_n
   + \omega'_n \psi'_{-n} \psi'_n
  ) ~.
\end{equation}
The zero mode contribution is given by
\begin{equation}
E_0 = \frac{\pi}{p^+} 
  \left( \left( \frac{p^{I_N}}{\pi} \right)^2 
       + \left( \frac{m}{3} \right)^2 (x^{i_N})^2
       + \left( \frac{m}{6} \right)^2 (x^{i'_N})^2
       + \frac{i}{\pi} \frac{m}{3} \psi_0 \gamma^4 \Omega \psi_0
       - \frac{i}{\pi} \frac{m}{6} \psi'_0 \gamma^4 \Omega \psi'_0
  \right) ~.
\end{equation}
In contrast to the closed string case, fermionic part has no ordering
ambiguity upon quantization and thus we leave it as intact.  For the
bosonic part, we introduce the creation and the annihilation operators,
\begin{eqnarray}
& & a^{i \dagger} = \sqrt{ \frac{3 \pi}{2m} }  
      \left( \frac{p^i}{\pi} + i \frac{m}{3} x^i \right)~,
 ~~~a^i = \sqrt{ \frac{3 \pi}{2m} }
      \left( \frac{p^i}{\pi} - i \frac{m}{3} x^i \right)~,
  \nonumber \\
& & a^{i' \dagger} = \sqrt{ \frac{6 \pi}{2m} }
      \left( \frac{p^{i'}}{\pi} + i \frac{m}{6} x^{i'} \right)~,
 ~~~a^{i'} = \sqrt{ \frac{6 \pi}{m} }
      \left( \frac{p^{i'}}{\pi} - i \frac{m}{6} x^{i'} \right) ~,
\end{eqnarray}
which give the usual commutation relations,
\begin{equation}
[ a^I, a^{J \dagger} ] = \delta^{IJ} ~.
\end{equation}
Then the zero mode contribution is finally read as
\begin{equation}
E_0 = \frac{m}{3 p^+} 
   \left(
      2 a^{i_N \dagger} a^{i_N} 
     + a^{i'_N \dagger} a^{i'_N}     
     + i \psi_0 \gamma^4 \Omega \psi_0 
     - \frac{i}{2} \psi'_0 \gamma^4 \Omega \psi'_0
     + e_0 
   \right) ~,
\label{ozm}
\end{equation}
where $e_0$ is the zero-point energy only coming from the bosonic
part
\begin{equation}
e_0 = n_N + \frac{1}{2} n'_N ~.
\label{zpe}
\end{equation}
Here $n_N$ ($n'_N$) is the number of Neumann directions in $1,2,3,4$
($5,6,7,8$) and constrained to be $n_N + n'_N = p-1$ for D$p$-brane.

\subsection{Supersymmetry of D-branes}

In this subsection we show that the D-branes found above are half-BPS
states of IIA superstring theory.  We would like to find the
supersymmetries preserved by the corresponding boundary conditions.
Firstly, we consider the boundary condition from the dynamical
supersymmetry transformation for bosonic string coordinates. For the
Dirichlet directions, we require that $\delta X^{I_D}$ vanish at the
open string boundaries, while for the Neumann directions, $\delta
\partial_\sigma X^{I_N}$ vanish at the boundaries.  Then using the
boundary conditions for the fermionic coordinates, (\ref{fbc}) and
noting that
\[
\partial_\sigma \psi^1_\pm \bigg|_{\sigma = 0,\pi}
= - \Omega \partial_\sigma \psi^2_\mp \bigg|_{\sigma=0, \pi}~,
\]
which is the reflection of the fact that the Euler-Lagrange equations
for $\psi^1$ and $\psi^2$ are related by $\sigma \leftrightarrow -
\sigma$ \cite{lam031}, we see that the supersymmetry variations both
for the Dirichlet and Neumann directions give rise to the same
condition which is
\begin{equation}
\gamma^I ( \Omega^T \epsilon^1_+ + \epsilon^2_-) = 0 ~.
\end{equation}
Therefore four dynamical supersymmetries generated by the
supersymmetry parameters with the relation
\begin{equation}
\epsilon^1_+ = - \Omega \epsilon^2_-~,
\label{dsrel}
\end{equation}
are preserved in the existence of the D-brane configurations given in
the previous subsection.  From the fermionic boundary condition,
(\ref{fbc}), we have the equation relating the dynamical supersymmetry
variations for the fermionic coordinates,
\begin{equation}
\delta \psi^1_\pm \bigg|_{\sigma = 0, \pi}
 = \Omega \delta \psi^2_\mp \bigg|_{\sigma=0, \pi} ~.
\end{equation}
If we insert the above relation (\ref{dsrel}) into this equation, we
have after some manipulation
\begin{equation}
\gamma^4 \Omega \gamma^4 \Omega = -1 ~,
\label{omega}
\end{equation}
which is nothing but the condition (\ref{oc}) in the previous
subsection that $\Omega$ should satisfy.  Thus, we see that the
condition for $\Omega$ is consistent with the dynamical supersymmetry.
 
For the kinematical supersymmetry transformations, from 
\begin{equation}
\tilde{\delta} \psi^1_\pm \bigg|_{\sigma =0, \pi}
= \Omega \tilde{\delta} \psi^2_\mp \bigg|_{\sigma=0,\pi} ~,
\label{ksfbc}
\end{equation}
one can easily see that the kinematical supersymmetry parameters
which satisfy 
\begin{equation}
\tilde{\epsilon}^1_\pm = \Omega \tilde{\epsilon}^2_\mp
\end{equation}
are compatible with the boundary conditions (\ref{ksfbc}), and
furthermore they give exactly the same condition for $\Omega$,
(\ref{oc}) or (\ref{omega}), and thus the existence of D-brane breaks
half the kinematical supersymmetry.  In total, D-brane listed in table
\ref{t2} preserves half the supersymmetry of the pp-wave: eight
kinematical and four dynamical.

\subsection{Comparison of IIA and M Theory Branes} 
It is interesting to compare D-branes found here with flat BPS states found 
in the context of matrix model in eleven-dimensional pp-wave background
\cite{hyu090}.
First of all, we do not expect to get the corresponding objects of 
D6 and D8 branes in the
matrix model set-up\cite{ban157}. 
Furthermore, all the D-branes found here is stretched along 
$x^-$-direction, as $x^-$-coordinate should satisfy Neumann boundary 
condition. As a result, D0-branes can not be seen in this formalism.
This may be regarded as the limitation of light-cone gauge fixed 
string theory. On the other hand, in the matrix model, only the
 membrane which is transverse to $x^-$ was identified, which 
also may be regarded 
as the limitation of matrix model. Henceforth, the D2-brane configuration 
found here has no counterpart in the matrix model and vice versa. 

On the 
other hand, the longitudinal M5-brane in the matrix model 
is stretched along
$x^-$-direction and would be identified with the D4-brane configuration 
found here. Here in IIA string theory side, 
we have two different configurations for flat D4-branes, 
depending on the direction they are stretched.      
The flat D4-branes stretched along four $x^{i'}$-coordinates can be 
identified with rotating longitudinal M5-brane found in \cite{hyu090}. Note 
that time-dependent rotation among coordinates with Neumann boundary 
conditions does not change the boundary conditions themselves. 

Other D4-brane solution which is stretched along two $x^i$-coordinates
and two $x^{i'}$-coordinates should also have counterpart 
in matrix model, though
not yet described in the literature\footnote{
This has been found independently by Jeong-Hyuck
Park and Sangheon Yi.}. The corresponding longitudinal five brane solutions 
in the matrix model is given as follows. Note that, in the matrix model, 
the relevant fermion transformation law is given by 
\begin{eqnarray}
\delta \theta  =
  \left(
    \sum_{I=1}^9  P^I \gamma^I
  + \frac{i}{2} \sum_{I=1}^9 [X^I, X^J] \gamma^{IJ}
  + \frac{\mu}{3R} \sum_{i=1}^3 X^i \Pi \gamma^i
  - \frac{\mu}{6R} \sum_{i'=4}^9 X^{i'} \gamma^{i'}\Pi
  \right) \eta~,
  \label{dyn}
\end{eqnarray}
where $P^I$ is the canonical conjugate momentum of bosonic coordinate $X^I$.
Consider the configurations
\begin{eqnarray}
X^{i}=x^i \cos(\mu t/3) + x^j \sin(\mu t/3) ~, \ \ \
X^{j}=-x^i\sin(\mu t/3) + x^j \cos(\mu t/3)~,\cr 
X^{i'}=x^{i'} \cos(\mu t/6) - x^{j'}\sin(\mu t/6) ~, \ \ \ 
X^{j'}=x^{i'} \sin(\mu t/6) + x^{j'}\cos(\mu t/6)~~,
\label{conf2}
\end{eqnarray}
while all other coordinates vanish. The rotating longitudinal five brane 
solution of second type, then, is given 
by time-independent $x^{i}$ and  $x^{i'}$ satisfying
the condition:
\begin{equation}
[x^{i}, x^{j} ] = \frac{1}{2}\epsilon_{ij i'j'}[x^{i'},
x^{j'}]~. \label{lc}
\end{equation}

\section*{Acknowledgments}
We would like to thank Jeong-Hyuck Park and Sangheon Yi for useful
discussions.  The work of S.H. was supported in part by Yonsei
University Research Fund of 2002-1-0222.  The work of H.S. was the
result of research activities (Astrophysical Research Center for the
Structure and Evolution of the Cosmos (ARCSEC)) supported by Korea
Science $\&$ Engineering Foundation.


\end{document}